\documentclass[preprint,aps]{revtex4}

\usepackage{graphicx}
\usepackage{dcolumn}
\usepackage{bm}
\usepackage{color}
\usepackage{amsmath}

\begin{document}
\title{Ideal cylindrical cloak: Perfect but sensitive to tiny perturbations}
\author{Zhichao Ruan\footnote{These authors contributed equally to this work.}$^{1,2}$,
Min Yan$^{*1}$, Curtis W. Neff$^{1}$, and
Min Qiu$^{1}$\footnote{Corresponding author. Electronic address: min@kth.se.}}

\affiliation{$^1$Laboratory of Optics, Photonics and Quantum
Electronics, Department of Microelectronics and Applied Physics,
Royal Institute of Technology (KTH), Electrum 229, 16440
Kista, Sweden\\
$^2$Joint Research Center of Photonics of the Royal Institute of
Technology (Sweden) and Zhejiang University, Zhejiang University,
Yu-Quan, 310027 Hangzhou, PR China}

\begin{abstract}
A cylindrical wave expansion method is developed to obtain the
scattering field for an ideal two-dimensional cylindrical
invisibility cloak. A near-ideal model of the invisibility cloak is
set up to solve the boundary problem at the inner boundary of the
cloak shell. We confirm that a cloak with the ideal material
parameters is a perfect invisibility cloak by systematically
studying the change of the scattering coefficients from the
near-ideal case to the ideal one. However, due to the slow
convergence of the zero$^{th}$ order scattering coefficients, a tiny
perturbation on the cloak would induce a noticeable field scattering
and penetration.
\end{abstract}

\maketitle

The exciting issue of exotic materials invisible to electromagnetic
(EM) waves was discussed in recent works
\cite{Pendry2006cef,leonhardt2006ocm,alu2005atp,miller2006pc,leonhardt2006nci,
cummer2006fws,schurig2006mec,Milton2006,Zolla2007,cai2007ocm,chen:241105}.
Based on a coordinate transformation of Maxwell's equations, Pendry
{\it{et al.}} first proposed an invisibility cloak, which can
protect objects inside the cloak from detection
\cite{Pendry2006cef}: When EM waves pass through the invisibility
cloak, the cloak will deflect the waves, guide them around the
object, and return them to the original propagation direction
without perturbing the exterior field. Numerical methods have been
applied to solve the EM problem involving invisibility cloaks
\cite{cummer2006fws,Zolla2007}, and an experimental result of the
invisibility cloak using metamaterial with simplified material
parameters has also recently been reported \cite{schurig2006mec}.
Yet, the ideal invisibility cloak has not been confirmed as a
perfect cloak, due to the extreme material parameters required (zero
or infinity) in the ideal cloak when approaching the inner boundary.
Also, numerical methods usually describe the material parameters
discretely, which can be computationally intensive in extreme cases.
Thus it is preferable to use an analytical or semi-analytical method
whenever possible.

In this paper, we will study the scattering for an ideal
invisibility cloak. We focus our analysis on the 2D cylindrical
cloak, because the wave equation can be simplified in comparison
with the 3D case, and a 2D invisibility cloak is more feasible to
fabricate \cite{schurig2006mec}. Here we take advantage of the
cylindrical geometry of the structure and use the cylindrical wave
expansion method to study the device semi-analytically. To avoid
extreme values (zeros or infinity) of material parameters at the
cloak's inner surface, we introduce a small perturbation into the
ideal cloak, and allow the perturbation to approach zero to study
the scattering problem for the ideal cloak. Such an asymptotic
analysis not only can confirm whether the ideal cloak would be
perfectly invisible or not, it also provides hints on how sensitive
such a device is to finite perturbations. A sensitivity analysis of
the invisibility cloak directly determines the possibility of its
application. Our studies show that the cylindrical invisibility
cloak is very sensitive to tiny perturbations of the material
parameters.

First, let's look at the wave equation inside a cylindrical cloak.
According to Ref. \cite{Pendry2006cef}, a simple transformation
\begin{equation}
\begin{array}{*{20}c}
   {r' = \frac{{b - a}}{b}r + a}, & {\theta ' = \theta }, & {z' = z}  \\
\end{array}
\end{equation}
can compress space from the cylindrical region $0 < r < b$ into the
annular region $a<r'<b$, where $a$ is the inner radius of the cloak,
$b$ is the outer radius of the cloak, and $r$, $\theta$ and $z$
($r'$, $\theta'$ and $z'$) are the radial, angular and vertical
coordinates in the original (transformed) system, respectively.
Following the approach in Ref.~\cite{Pendry2006cef}, the
permittivity and permeability tensor components for the cloak shell
can be given as
\begin{equation}
\begin{array}{*{20}c}
   {\varepsilon _r  = \mu _r  = \frac{{r - a}}{r}}, & {\varepsilon _\theta   = \mu _\theta   = \frac{r}{{r - a}}},  \\
   {\varepsilon _z  = \mu _z  = \left( {\frac{b}{{b - a}}} \right)^2 \frac{{r - a}}{r}}, & {}  \label{eq:paramater}\\
\end{array}
\end{equation}
and air is assumed for the ambient environment and the interior
regions. In the following, the transverse-electric (TE) polarized
electromagnetic field is considered (i.e. the electrical field only
exists in the $z$-direction) , however the transverse-magnetic
derivation follows in similar manner. Throughout the paper, a
$\exp(-i\omega t)$ time dependence is assumed. For the TE-polarized
wave, only $\varepsilon _z$, $\mu _r $, and $ \mu _\theta$ are
relevant to the following general wave equation governing the $E_z$
field in the cloak's cylindrical coordinate
\begin{equation}
\frac{1}{{\varepsilon _z r}}\frac{\partial }{{\partial
r}}(\frac{r}{{\mu _\theta  }}\frac{{\partial E_z }}{{\partial r}}) +
\frac{1}{{\varepsilon _z r^2 }}\frac{\partial }{{\partial \theta
}}(\frac{1}{{\mu _r }}\frac{{\partial E_z }}{{\partial \theta }}) +
k_0^2 E_z  = 0, \label{eq:WaveEqMaster}
\end{equation}
where $k_0$ is the wave vector of light in vacuum. If we substitute
Eq.~\ref{eq:paramater} for $\varepsilon _z$, $\mu _r $, and $ \mu
_\theta$, we find
\begin{equation}
r^2 \frac{{\partial ^2 E_z }}{{\partial r^2 }} + r\mu _\theta
\frac{{\partial E_z }}{{\partial r}} + \varepsilon _z \mu _\theta
r^2 k_0^2 E_z  + \frac{{\mu _\theta  }}{{\mu _r }}\frac{{\partial ^2
E_z }}{{\partial \theta ^2 }} = 0 \label{eq:WaveCloak}
\end{equation}
Equation \ref{eq:WaveCloak}can be solved by a separation of
variables $E_z  = \Psi (r)\Theta (\theta )$ and the introduction of
a constant $l$:
\begin{equation}
 (r - a)^2 \frac{{\partial ^2 \Psi }}{{\partial r^2 }} + (r - a)\frac{{\partial \Psi }}{{\partial r}} + [(\frac{b}{{b - a}})^2 (r - a)^2 k_0^2  - l^2 ]\Psi  = 0 \\
\label{eq:WaveCloakSplitterPsi}
\end{equation}
\begin{equation}
 \frac{{\partial ^2 \Theta }}{{\partial \theta ^2 }} + l^2 \Theta  = 0. \\
\label{eq:WaveCloakSplitterTheta}
\end{equation}
Equation \ref{eq:WaveCloakSplitterPsi} is the $l^{th}$-order Bessel
differential equation, and the general solution of
Eq.~\ref{eq:WaveCloakSplitterTheta} is $\exp (il\theta )$.
Therefore, there exists a simple set of solutions to $E_z$ in the
cloak shell of the form
\begin{equation}
F_l (k_1 (r-a))\exp (il\theta ) \label{eq:WaveCloakSolution}
\end{equation}
where $k_1= k_0 b/(b-a)$, $F_l$ is the $l$-order Bessel function,
and $l$ is an integer number as required by the rotational boundary
condition.


Let us consider the scattering problem in which an arbitrary wave is
incident on the cloak. According to the rigorous scattering theory
\cite{Hulst2007}, the incident field in the 2D case can be expanded
in the cloak's coordinates with the following expression
\begin{equation}
E_z^{in}  = \sum\limits_l {\alpha _l^{in} J_l (k_0 r)\exp (il\theta
)}, \label{eq:Expansion}
\end{equation}
where $J_l$ is the $l^{th}$-order Bessel function of the first kind.
The scattering field can also be expanded as
\begin{equation}
E_z^{sc}  = \sum\limits_l {\alpha _l^{sc} H_l (k_0 r)\exp (il\theta
)},
\end{equation}
where $H_l$ is the $l^{th}$-order Hankel function of the first kind.

\begin{figure}[h]
\centerline{\includegraphics[width=3.0in]{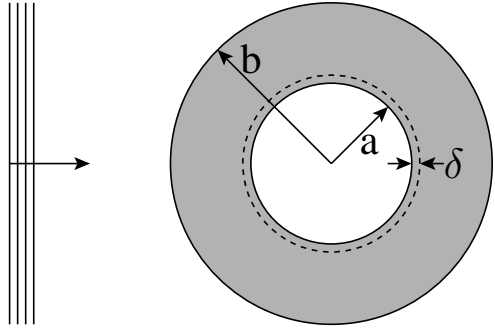}}
\caption{\label{fig:schematic} The schematic of a near-ideal
invisibility cloak: The distribution of the material parameter is
the same as the ideal one shown in Eq.~\ref{eq:paramater}, and the
outer boundary is still fixed at $r=b$. However, the actual inner
boundary is at $r=a+\delta$, where $\delta $ is a very small
positive number. }
\end{figure}

We note that the scattering coefficients cannot be directly obtained
for the ideal cloak
 since $\varepsilon _z \to 0$, $\mu _r \to 0 $, and $\mu _\theta \to \infty
$ when $r \to a$, and the Bessel function of the second kind in
Eq.~\ref{eq:WaveCloakSolution} has a singularity at $r=a$. In order
to circumvent this, we introduce a small perturbation to the ideal
cloak which we refer to as the near-ideal cloak, see
Fig.~\ref{fig:schematic}. We expand the inner boundary of the cloak
shell slightly, so that it is located at $r=a+\delta $, where
$\delta $ is a very small positive number. However, the material
parameters are still calculated according to Eq.~\ref{eq:paramater}
as if the inner boundary is unchanged. The outer boundary remains
fixed at $r=b$. When $\delta \to 0$, our model will be equivalent to
the ideal cloak. Now the electric-field in each region can be given
by
\begin{equation}
\begin{array}{*{20}c}
   {(b < r)} \hfill & {E_z  = } \hfill & {\sum\limits_l {\alpha _l^{in} J_l (k_0 r)\exp (il\theta ) + \alpha _l^{sc} H_l (k_0 r)\exp (il\theta )} } \hfill  \\
   {(a + \delta  < r < b)} \hfill & {E_z  = } \hfill & {\sum\limits_l {\alpha _l^1 J_l (k_1 (r - a))\exp (il\theta ) + \alpha _l^2 H_l (k_1 (r - a))\exp (il\theta )} } \hfill  \\
   {(r < a + \delta )} \hfill & {E_z  = } \hfill & {\sum\limits_l {\alpha _l^3 J_l (k_0 r)\exp (il\theta )} } \hfill
\label{eq:EachLayerField}
\end{array}
\end{equation}
where $ \alpha _l^{i} (i=1,2,3)$ are the expansion coefficients for
the resulting field inside the cloak.

The tangential fields $E_z$ and $H_\theta$ (which can be obtained
from $E_z$), should be continuous across the interfaces at
$r=a+\delta$ and $r=b$; and the orthogonality of $\exp(il\theta)$
allows waves in each Bessel order to decouple. Thus, we can have the
following four equations:
\begin{subequations}
\begin{equation}
 \alpha _l^{in} J_l (k_0 b) + \alpha _l^{sc} H_l (k_0 b) = \alpha _l^1 J_l (k_1 (b - a)) + \alpha _l^2 H_l (k_1 (b - a)) \\
\label{eq:BoundaryEqa}
\end{equation}
\begin{equation}
 \alpha _l^1 J_l (k_1 \delta ) + \alpha _l^2 H_l (k_1 \delta ) = \alpha _l^3 J_l (k_0 (a + \delta )) \\
 \label{eq:BoundaryEqb}
\end{equation}
\begin{equation}
 k_0 \alpha _l^{in} J_l '(k_0 b) + k_0 \alpha _l^{sc} H_l '(k_0 b) = \frac{{k_1 }}{{\mu _\theta  (b)}}\alpha _l^1 J_l '(k_1 (b-a)) + \frac{{k_1 }}{{\mu _\theta  (b)}}\alpha _l^2 H_l '(k_1 (b-a)) \\
 \label{eq:BoundaryEqc}
\end{equation}
\begin{equation}
 \frac{{k_1 }}{{\mu _\theta  (a + \delta )}}\alpha _l^1 J_l '(k_1 \delta ) + \frac{{k_1 }}{{\mu _\theta  (a + \delta )}}\alpha _l^2 H_l '(k_1 \delta ) = k_0 \alpha _l^{3} J_l '(k_0 (a + \delta )) \\
\label{eq:BoundaryEqd}
\end{equation}
\label{eq:BoundaryEq}
\end{subequations}
which is a set of linear equations. Thus each order expansion
coefficient in each material region can be exactly solved. In turn
we can obtain the fields in each region.

As a direct result of this set linear equations, we can prove that
when $\delta \to 0$, $\alpha _l^{sc} = \alpha _l^{2} \to 0$ ,
$\alpha _l^{1} = \alpha _l^{in}$, and $\alpha _l^{3} \to 0$ for any
$\alpha _l^{in}$, i.e., the ideal cloak is a perfect invisibility
cloak. Firstly, it can be assumed that $\left| \alpha _l^{i}
(i=sc,1,2,3) \right|$ must be finite. Otherwise, the scattering
field would be infinite if the incident field has the $l^{th}$ order
component. Secondly, due to $ k_1 (b-a)=k_0 b$ and $k1=k_0 {\mu
_\theta  (b)}$, when $\delta \to 0$, Eq.~\ref{eq:BoundaryEqa} and
\ref{eq:BoundaryEqc} become $ (\alpha _l^{in}-\alpha _l^1 ) J_l (k_0
b) +( \alpha _l^{sc}- \alpha _l^2 ) H_l (k_0 b) = 0$ and $ (\alpha
_l^{in}-\alpha _l^1 ) J'_l (k_0 b) +( \alpha _l^{sc}- \alpha _l^2 )
H'_l (k_0 b) = 0$, respectively. Since $b$ can be arbitrary and the
Bessel functions are not always zeros, $\alpha _l^{in}=\alpha _l^1 $
and $ \alpha _l^{sc}=\alpha _l^2$ must be satisfied. Thirdly, from
Eq.~\ref{eq:BoundaryEqb}, we can obtain the following inequality
\begin{equation}
\left| \alpha _l^2 H_l (k_1 \delta ) \right| \leq \left| \alpha _l^3
J_l (k_0 (a + \delta )) \right| +  \left| \alpha _l^1 J_l (k_1
\delta ) \right| .
\end{equation}
When $\delta \to 0$, the right side of the above inequality
approaches a finite value but $\left| H_l (k_1 \delta ) \right|$
approaches infinity on the left side. Thus, $\left| \alpha _l^2
\right| $ must approach zero. Finally, from
Eq.~\ref{eq:BoundaryEqb}, we can also obtain that $\left| \alpha
_l^2 H'_l (k_1 \delta ) \right| \leq \left| \alpha _l^3
\frac{k_0}{k_1 } J'_l (k_0 (a + \delta )) \right| + \left|  \alpha
_l^1 J'_l (k_1 \delta ) \right| $. While from
Eq.~\ref{eq:BoundaryEqd}, we have
\begin{equation}
\left| k_0 \alpha _l^{3} J_l '(k_0 (a + \delta )) \right| \leq
\left| \frac{{k_1 }}{{\mu _\theta (a + \delta )}}\alpha _l^1 J_l
'(k_1 \delta ) \right| + \left| \frac{{k_1 }}{{\mu _\theta  (a +
\delta )}}\alpha _l^2 H_l '(k_1 \delta ) \right|.
\end{equation}
Since ${\mu _\theta (a + \delta )} \to \infty$ and the right side of
the above inequality approaches zero when $\delta \to 0$, we obtain
that $\left|\alpha _l^{3}\right| \to 0$. Consequently, this argument
proves that the scattering field and the field in the interior
region of the cloak are zero when $\delta=0$, i.e., the ideal cloak
is a perfect invisibility cloak.

\begin{figure}[h]
\includegraphics[width=3.4in]{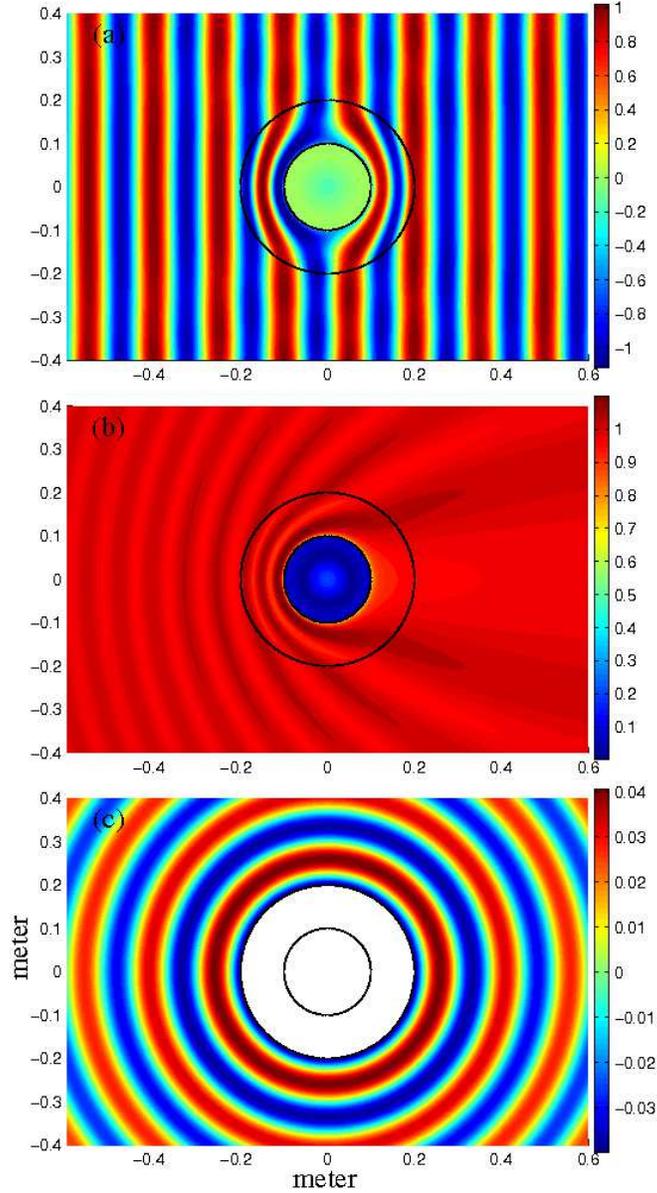} \hfill
\caption{\label{fig:PlaneWave} (Color online) (a) Snapshot of the
resulting electric-field distribution, (b) the corresponding norm in
the vicinity of the cloaked object, and (c) the snapshot of the
corresponding scattering field outside the cloak for the near-ideal
cloak with $\delta =10^{-5}a$ and when a plane wave is
perpendicularly incident on the cloak. The black lines outline the
cloak shell. Axis unit: meter.}
\end{figure}

Although we have just confirmed that the ideal cloak can provide
perfect invisibility, further study the near-ideal cloak by the
above analytical method illuminates how sensitive the parameter
$\delta$ is to the performance of the cloak. As an example, we use
the same material parameters in Ref.~\cite{cummer2006fws} where the
inner radius of the cloak is $a=0.1$m, the outer radius of the cloak
is $b=0.2$m, and the frequency of the incident plane wave is 2GHz.
Similarly, we also consider a plane wave incident on the cloak,
where the expansion coefficients in Eq.~\ref{eq:Expansion} are
\begin{equation}
\alpha _l^{in}  = i^l A \exp ( - ik_0 r_1 \cos (\varphi  + \theta _1
) - il\varphi ), \label{eq:PlaneWave}
\end{equation}
where $(r_1,\theta_1,0)$ is the coordinate of the phase reference
point, $A$ is the amplitude of the plane wave, and $\varphi $ is the
incident angle \cite{Felbacq1994}. Here the phase reference point is
set at $r_1=4a$ and $\theta_1=\pi$, the amplitude is $A=1$, and the
incident angle is $\varphi =0$ (i.e. the plane wave propagates from
left to right). We use 31 Bessel terms ($- 15 \le l \le 15$) to
calculate the scattering field for the near-ideal cloak with $\delta
=10^{-5}a$. The number of expansion terms is sufficient  for
convergence of the calculated fields. Figure~\ref{fig:PlaneWave}
shows the snapshot of the resulting electric-field distribution
(i.e. the real part of the electric-field phasor), and the
corresponding norm in the vicinity of the cloaked object. The
electric-field distribution clearly demonstrates the cloaking effect
of the near-ideal cloak to the incident plane wave. However, the
norm of the electric-field (Fig.~\ref{fig:PlaneWave} (b)) reveals
that there is still a little bit of the field in the cloak interior
and an obvious scattering ripple around the cloak. The amplitude of
the resulting electric-field at the center is $0.197$.
The snapshot of the scattering field
(Fig.~\ref{fig:PlaneWave} (c)) shows that it propagates almost
isotropically in all angles. Even though the amplitude of the
scattering field is much smaller than that of the incident plane
wave, the interference of the incident plane wave and the scattering
field creates the ripples in the norm (Fig.~\ref{fig:PlaneWave}
(b)).

\begin{figure}[t]
\includegraphics[width=3.4in]{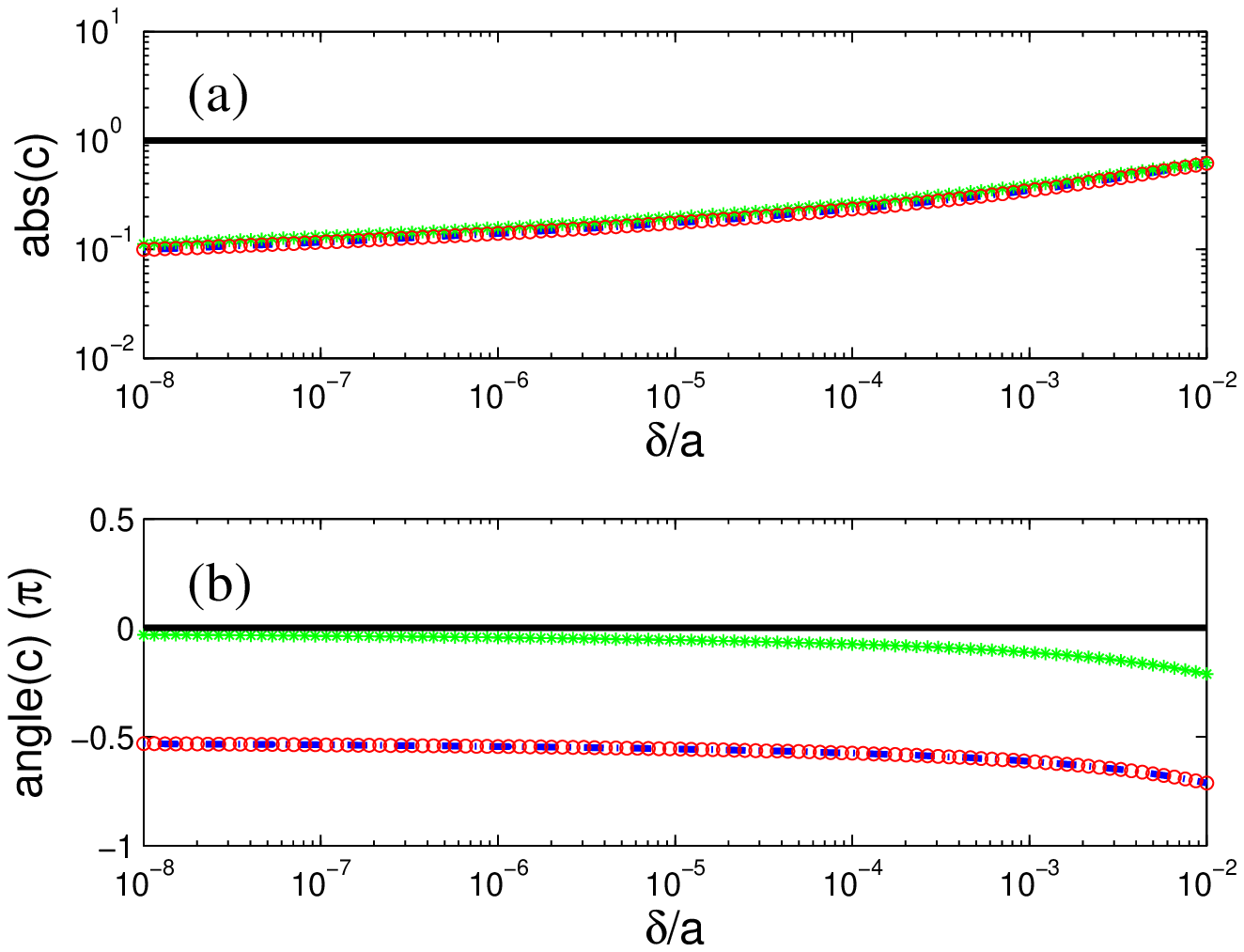} 
\includegraphics[width=3.4in]{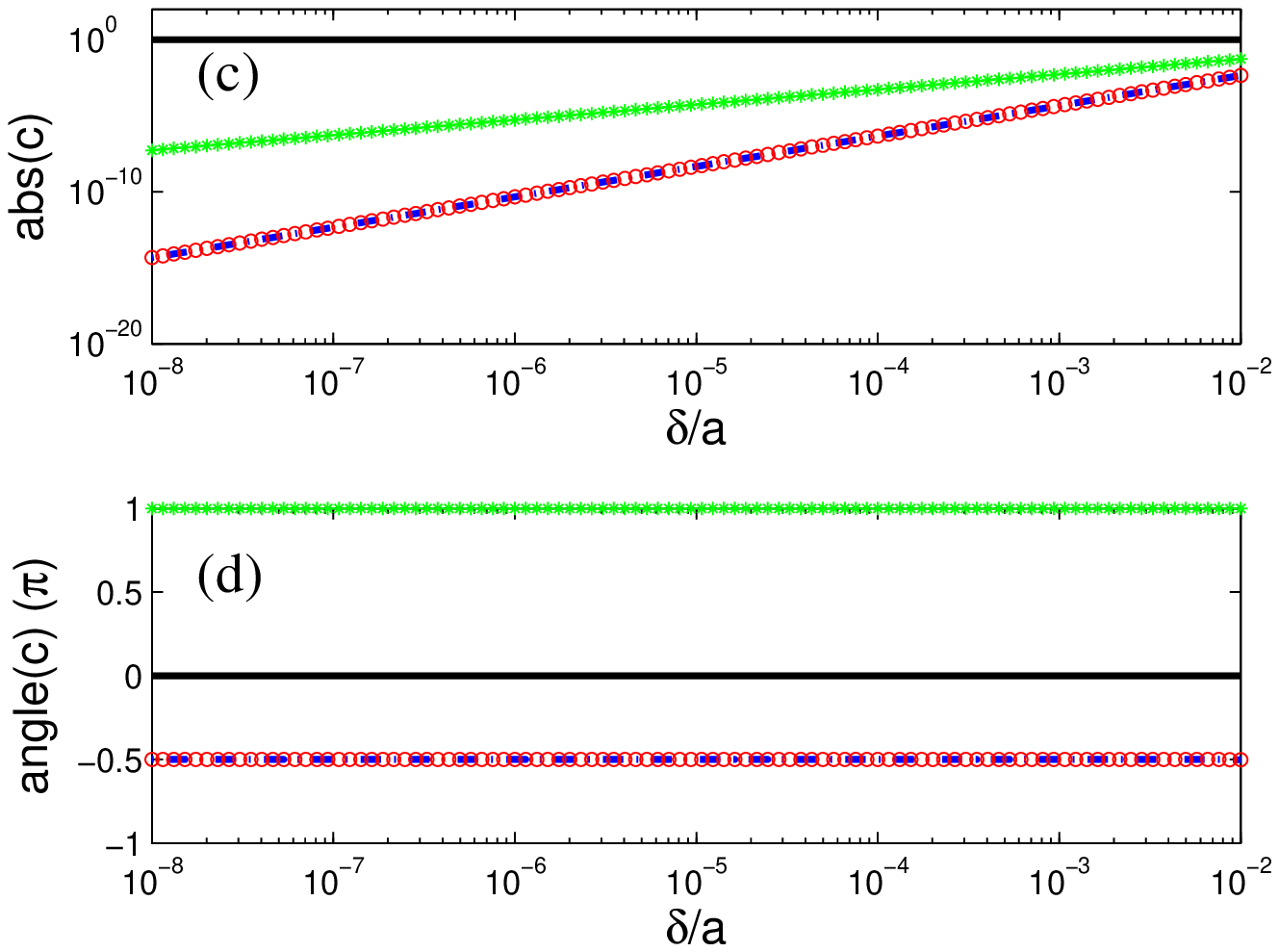} 
\caption{\label{fig:ScatteringCoeff} (Color online) The amplitude
and phase of the scattering coefficients for the different $\delta$,
where (a)-(b) and (c)-(d) correspond to the cases of $l=0$ and $l=1$
respectively. $c_l^{sc}$ is denoted by the blue point-dashed line,
$c_l^{(1)}$ (the solid black line), $c_l ^{(2)}$ (the red
circle-marked), and $c_l ^{(3)}$ (the green star-marked). }
\end{figure}

Since each order expansion coefficient of the scattering field is
only relevant to each order expansion coefficient of the incident
field (cf. Eq.~\ref{eq:BoundaryEq}), we can define the scattering
coefficient for each order as
\begin{equation}
 c_l^{sc}  = \frac{\alpha _l^{sc}}{\alpha _l^{in}}.
\end{equation}
These coefficients for the field inside the cloak $c_l^{(i)} =
\alpha _l^{i} /\alpha _l^{in}, i=1,2,3$ can also be defined in the
same way. To study the ideal cloak, we more $\delta$ closer to $0$.
The amplitude and the phase of these coefficients for $
10^{-8}a<\delta<10^{-2}a$ are shown in
Fig.~\ref{fig:ScatteringCoeff}, where (a)-(b) and (c)-(d) correspond
to the cases of $l=0$ and $l=1$ , respectively.

From Fig.~\ref{fig:ScatteringCoeff}, it is clear that $c_l^{(1)}$ is
always equal to $1$ for both cases. That is, the incident field
propagates into the cloak without any reflection at the outer
boundary, which coincides with the explanation of the cloaking
effect from the coordination transformation approach
\cite{Pendry2006cef}. The same behavior for the scattering fields
occurs at the outer boundary, where they propagate from inside to
outside without any reflection, thus $c_l ^{sc}$ is always equal to
$c_l ^{(2)}$.

Our computational results also confirm that both $c_l ^{sc}$ and
$c_l ^{(3)}$ approach zero when $\delta \to 0$. In particular,
compared with the case of $l=0$, $\left| { c_l ^{sc}} \right| $ and
$\left| { c_l ^{(3)} } \right| $ for $l=1$ are much smaller, and
approach zero more rapidly. This is also observed for the other
higher order cases. Thus, in the case of the plane wave incident,
where $\left| {\alpha _l^{in} } \right|$ is the same for each order,
the dominating term of the scattering field outside of the cloak is
of the form of the zero$^{th}$-order Hankel function of the first
kind. Meanwhile, the resulting field in the interior region has a
dominating term of $J_0(k_0r)$ . This explains the near azimuthally
invariable distribution of the field in the interior region (see
Fig.~\ref{fig:PlaneWave}(a)) and the scattering field outside the
cloak (see Fig.~\ref{fig:PlaneWave}(c)), which has also been
mentioned in Ref.~\cite{cummer2006fws}.

It is worth noting that the zero$^{th}$ order scattering
coefficients $c_0^{sc} $ and $c_0^{(3)} $ decrease extremely slowly
with reduced $\delta$, e.g. when $\delta$ is decreased from
$10^{-5}a$ to $10^{-8}a$, $\left| {c_0^{sc} } \right|$ decreased
only from $0.175$ to $0.099$. By utilizing the arbitrary calculation
precision of the software MATHEMATICA, we found that the convergence
of the limit is so slow that even for $\delta=10^{-99}a$ (i.e.
$\varepsilon _z \approx 4 \times 10^{-99}$, $\mu_r \approx
10^{-99}$, and ${\mu _\theta } \approx 10^{99}$ at the inner
boundary in this case), $\left| {c_0^{sc} } \right|=6.973\times
10^{-3}$. Therefore, we conclude that even though an cloak with the
ideal material parameters in Ref.~\cite{Pendry2006cef} is a perfect
cloak, a non-ideal invisibility cloak does not provide a good enough
cloaking effect due to the slow convergence of $\left| {c_0^{sc} }
\right|$ and $\left| {c_0^{(0)} } \right|$.

In conclusion, we have used the cylindrical wave expansion method to
study the electromagnetic scattering properties of a 2D invisibility
cloak. A near-ideal model of the invisibility cloak is set up to
solve the boundary problem at the inner boundary of the cloak shell.
By systematically studying the change of the scattering coefficients
from the near-ideal case to the ideal one, we confirm that the cloak
with the ideal material parameter is a perfect invisibility cloak.
But due to the slow convergence of the scattering coefficients, a
tiny perturbation on the cloak would induce a noticeable field
scattering and penetration. We also proved that the scattered and
penetrated fields are dominated by zero$^{th}$-order cylindrical
waves. Though our work has focused on the 2D cylindrical cloak, it
can be reliably extended to the 3D spherical case. Our method and
results are also useful for either designing or detecting this type
of the invisibility cloak.

This work is supported by the Swedish Foundation for Strategic
Research (SSF) through the INGVAR program, the SSF Strategic
Research Center in Photonics, and the Swedish Research Council (VR).
Z.C.R. acknowledges the partial support from the National Basic
Research Program (973) of China under Project No. 2004CB719800.


\end{document}